\documentclass[twoside]{IEEEtran}
\PassOptionsToPackage{ruled, linesnumbered}{algorithm2e}
\usepackage[latin9]{inputenc}
\usepackage{color}
\usepackage{mathrsfs}
\usepackage{algorithm2e}
\usepackage{amsmath}
\usepackage{amssymb}
\usepackage{graphicx}

\makeatletter
\usepackage{balance}
\usepackage{pgfplots}
\usepackage{pgfplotstable}
\usepackage{tikz}
\usetikzlibrary{calc}
\usepackage{cite}

\newcommand{\herm}{^{{\dagger}}}
\newcommand{\trans}{^{\mathsf{T}}}

\DeclareMathOperator{\tr}{Tr}
\DeclareMathOperator{\vecd}{\mathrm{vec_d}}

\DeclareMathOperator{\diag}{diag}
\DeclareMathOperator{\maximize}{maximize}
\DeclareMathOperator{\st}{subject~to}
\DeclareMathOperator{\argmin}{argmin}

\makeatother

\begin{document}
\title{\LARGE{Achievable Rate Maximization for Underlay Spectrum Sharing
MIMO System with Intelligent Reflecting Surface}}
\author{Vaibhav Kumar, \textit{Member, IEEE,} Mark F. Flanagan, \textit{Senior
Member, IEEE,} Rui Zhang, \textit{Fellow, IEEE,}\\
and Le-Nam Tran, \textit{Senior Member, IEEE}\thanks{This publication has emanated from research conducted with the financial
support of Science Foundation Ireland (SFI) and is co-funded under
the European Regional Development Fund under Grant Number 17/CDA/4786.\protect \\
V. Kumar, M. F. Flanagan, and L.-N. Tran are with School of Electrical
and Electronic Engineering, University College Dublin, D04 V1W8 Dublin,
Ireland (email: vaibhav.kumar@ieee.org; mark.flanagan@ieee.org; nam.tran@ucd.ie).\protect \\
R. Zhang is with the Department of Electrical and Computer Engineering,
National University of Singapore, Singapore (e-mail: elezhang@nus.edu.sg).}}
\maketitle
\begin{abstract}
\textcolor{black}{In this letter, the achievable rate maximization
problem is considered for intelligent reflecting surface (IRS) assisted
multiple-input multiple-output (MIMO) systems in an underlay spectrum
sharing scenario, subject to interference power constraints at the
primary users. The formulated non-convex optimization problem is challenging
to solve due to its non-convexity as well as coupling design variables
in the constraints. Different from existing works that are mostly
based on alternating optimization (AO), we propose a penalty }dual
decomposition based gradient projection (PDDGP) algorithm to solve
this problem. We also provide a convergence proof and a complexity
analysis for the proposed algorithm. We benchmark the proposed algorithm
against two known solutions, namely a minimum mean-square error based
AO algorithm and an inner approximation method with block coordinate
descent. Specifically, the complexity of the proposed algorithm \textcolor{black}{grows
}\textit{\textcolor{black}{linearly}}\textcolor{black}{{} with respect
to the number of reflecting elements at the IRS,} while that of the
two benchmark methods\textcolor{black}{{} grows with the third power
of the number of IRS elements.} Moreover, numerical results show that
the proposed PDDGP algorithm yields considerably higher achievable
rate than the benchmark solutions.
\end{abstract}


\section{Introduction}

With the recent development of software-controlled metasurface technology,
intelligent reflecting surfaces (IRSs) are being regarded as a promising
technology to enhance the performance of next-generation wireless
communications~\cite{IRS_Tutorial}. By virtue of flexible configuration
of the radio propagation environment, IRSs are capable of increasing
the spectral as well as the energy efficiency of wireless communication
systems. On the other hand, spectrum sharing has great potential to
support massive connectivity in the congested radio spectrum for beyond-fifth-generation
(B5G) communications standards~\cite{SS_Mag,TVT_NOMA}. Therefore,
the design and analysis of IRS-assisted spectrum sharing to enhance
the achievable rate of wireless communication systems have gained
considerable attention in the recent years.

In this context, the problem of optimal active and passive beamforming
for secondary user (SU) rate maximization in an IRS-assisted single-input
single-output (SISO) and multiple-input single-output (MISO) spectrum
sharing systems were considered in~\cite{ZhangCL,SchoberSPAWC,Larsson_MISO_TCOM,Robust_MISO_TCCN,VerticalBeamforming_WCL}.
However, studies on SU rate maximization for the more challenging
IRS-assisted multiple-input multiple-output (MIMO) underlay spectrum
sharing systems is still limited. In order to maximize the weighted
sum rate of the secondary network in an IRS-assisted MIMO system in
an underlay spectrum sharing scenario with multiple secondary receivers
(SRs) and a single primary receiver (PR), a block coordinate descent
algorithm along with an inner approximation method (BCD-IAM) was proposed
in~\cite{MIMO_BCD_TVT}. Similarly, in~\cite{TWC_Ding}, the authors
considered the problem of weighted sum rate maximization of the secondary
network for an IRS-assisted MIMO spectrum sharing system consisting
of multiple SRs as well as multiple PRs, where they adopted a weighted
minimum mean-square error (WMMSE) method and an alternating optimization
(AO)-based algorithm to find a suboptimal solution. On the other hand,
a suboptimal solution to the problem of achievable secrecy rate maximization
for an IRS-assisted MIMO system using AO, barrier method, SDR and
minorization-maximization (MM), was presented in~\cite{Limeng_WCNC}.

It is noteworthy that the methods proposed to maximize the SU rate
for MISO systems in~\cite{ZhangCL,SchoberSPAWC,Larsson_MISO_TCOM,Robust_MISO_TCCN,VerticalBeamforming_WCL}
are not applicable to MIMO systems, unless each receive antenna in
the MIMO system is treated as a separate user. However, this reduces
the beamforming/multiplexing gains which are major benefits of MIMO
systems. The method presented in~\cite{MIMO_BCD_TVT} for MIMO systems
is proposed only for the case of a single PR. Similarly, the system
considered in~\cite{Limeng_WCNC}, when omitting the eavesdropper,
reduces to the conventional IRS-assisted MIMO spectrum sharing system.
However, the solution proposed in~\cite{Limeng_WCNC} is limited
to the scenario where the direct links between the (secondary) transmitter
and (primary and secondary) receivers are absent. On the other hand,
the AO-based solution proposed in~\cite{TWC_Ding} can be inefficient
due to the possibly high coupling of design variables, especially
when the SU rate performance is dominated by the interference constraints
at the PRs. Motivated by the drawbacks of existing methods, in this
letter we propose a low-complexity but efficient algorithm to maximize
the achievable rate of the SU in an IRS-assisted MIMO underlay spectrum
sharing system consisting of multiple PRs, where direct links between
the (secondary) transmitter and (primary and secondary) receivers
are also present. Our main contributions in this letter are as follows:
(i) we propose a penalty dual decomposition based gradient projection
(PDDGP) algorithm to maximize the achievable rate; (ii) we provide
a convergence proof and complexity analysis of the proposed algorithm;
and (iii) we provide extensive numerical results to demonstrate the
superiority of the proposed solution over the benchmark solutions
provided in~\cite{MIMO_BCD_TVT} and~\cite{TWC_Ding}, and to show
the effect of different system parameters on the achievable rate.

\paragraph*{Notations}

Bold lowercase and uppercase letters denote vectors and matrices,
respectively. $\mathbf{X}^{*}$, $\mathbf{X}\trans$, $\mathbf{X}\herm$,
$\left|\mathbf{X}\right|$, $\left\Vert \mathbf{X}\right\Vert $ and
$\tr(\mathbf{X})$ denote the complex conjugate, (ordinary) transpose,
Hermitian transpose, determinant, Frobenius norm and trace of a matrix
$\mathbf{X}$, respectively. The vector space of complex matrices
of size $M\times N$ is denoted by $\mathbb{C}^{M\times N}$. $\diag(\mathbf{x})$
denotes a square diagonal matrix where the elements of $\mathbf{x}$
comprise the main diagonal, $\vecd(\mathbf{X})$ denotes the column
vector containing the elements of the main diagonal of the matrix
$\mathbf{X}$, and $\nabla_{\mathbf{X}}f(\cdot)$ denotes the complex-valued
gradient of $f(\cdot)$ with respect to (w.r.t.) $\mathbf{X}^{*}$.
$\mathbb{E}\{\cdot\}$ denotes the expectation operator. $\mathbf{X}\succeq\mathbf{Y}$
represents that $\mathbf{X}-\mathbf{Y}$ is positive semidefinite,
\textcolor{black}{and $\Pi_{\mathcal{X}}(\mathbf{x})$ denotes th}e
Euclidean projection of the point $\mathbf{x}$ onto the set $\mathcal{X}$,
i.e., $\Pi_{\mathcal{X}}(\mathbf{x})\triangleq\argmin_{\hat{\mathbf{x}}\in\mathcal{X}}\left\Vert \hat{\mathbf{x}}-\mathbf{x}\right\Vert $.

\begin{figure}[t]
\begin{centering}
\includegraphics[scale=0.45]{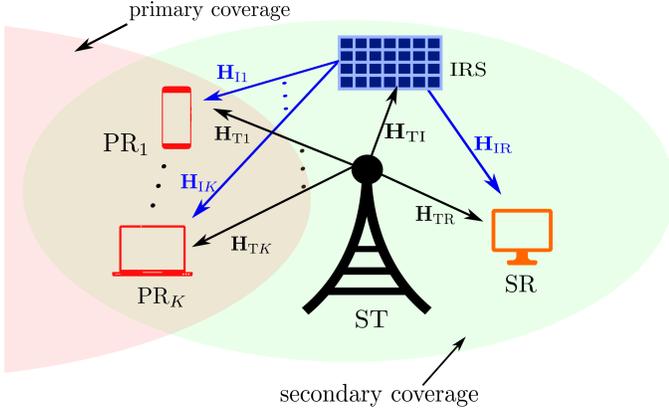}
\par\end{centering}
\caption{System model for IRS-assisted MIMO underlay spectrum sharing.}
\label{fig:SysMod}
\end{figure}

\begin{algorithm}[t]
\caption{Gradient projection algorithm to solve~\eqref{eq:OptProbTransformed}
for fixed $\boldsymbol{\upsilon}$ and $\rho$.}

\label{algoGP}

\KwIn{ $\boldsymbol{\theta}_{0}$, $\mathbf{X}_{0}$, $\mathbf{s}_{0}$,
$\boldsymbol{\upsilon}$, $\mu_{0}$, $\alpha_{0}$, $\rho$ }

\KwOut{ $\boldsymbol{\theta}_{n},\mathbf{X}_{n}$}

$n\leftarrow1$

\Repeat{convergence }{

$\!\!\boldsymbol{\theta}_{n}=\Pi_{\vartheta}(\boldsymbol{\theta}_{n\!-\!1}\!+\!\mu_{n}\nabla_{\boldsymbol{\theta}}\hat{R}_{\boldsymbol{\upsilon},\rho}(\mathbf{X}_{n\!-\!1},\boldsymbol{\theta}_{n\!-\!1},\mathbf{s}_{n\!-\!1}))$\;

$\!\!\mathbf{X}_{n}=\Pi_{\mathcal{X}}(\mathbf{X}_{n\!-\!1}+\alpha_{n}\nabla_{\mathbf{X}}\hat{R}_{\boldsymbol{\upsilon},\rho}(\mathbf{X}_{n\!-\!1},\boldsymbol{\theta}_{n},\mathbf{s}_{n\!-\!1}))$\;

$\!\!s_{n,k}=\max\{0,P_{k}\!-\!\tr(\mathbf{Z}_{k}\mathbf{X}_{n}\mathbf{Z}_{k}\herm)\}$,$\forall k\in\mathcal{K}$

$\!\!$where $\mathbf{Z}_{k}=\mathbf{H}_{\mathrm{I}k}\boldsymbol{\Theta}_{n}\mathbf{H}_{\mathrm{TI}}+\mathbf{H}_{\mathrm{T}k}$\;

$\!\!$$n\leftarrow n+1$\;

}
\end{algorithm}

\section{System Model and Problem Formulation}

Let us first consider an IRS-assisted MIMO system in an underlay spectrum
sharing scenario shown in Fig.~\ref{fig:SysMod}, consisting of one
secondary transmitter (ST), one SR, and $K$ PRs. The number of antennas
at the ST and SR are denoted by $N_{\mathrm{T}}$ and $N_{\mathrm{R}}$,
respectively, while the $K$ PRs are each equipped with $N_{\mathrm{P}}$
antennas. The communication between the ST and SR is assisted by an
IRS consisting of $N_{\mathrm{I}}$ low-cost passive reflecting elements.
The ST-SR, ST-IRS, and IRS-SR channels \textcolor{black}{are denoted
by} $\mathbf{H}_{\mathrm{TR}}\in\mathbb{C}^{N_{\mathrm{R}}\times N_{\mathrm{T}}}$,
$\mathbf{H}_{\mathrm{TI}}\in\mathbb{C}^{N_{\mathrm{I}}\times N_{\mathrm{T}}}$,
and $\mathbf{H}_{\mathrm{IR}}\in\mathbb{C}^{N_{\mathrm{R}}\times N_{\mathrm{I}}}$,
respectively. On the other hand, the channel between the ST and the
$k$-th PR ($k\in\mathcal{K}\triangleq\left\{ 1,2,\ldots,K\right\} $)
is denoted by $\mathbf{H}_{\mathrm{T}k}\in\mathbb{C}^{N_{\mathrm{P}}\times N_{\mathrm{T}}}$,
while the channel between the IRS and the $k$-th PR is denoted by
$\mathbf{H}_{\mathrm{I}k}\in\mathbb{C}^{N_{\mathrm{P}}\times N_{\mathrm{I}}}$.
Let $\boldsymbol{\theta}\triangleq[\theta_{1}\ \theta_{2}\ \ldots\ \theta_{N_{\mathrm{I}}}]\trans\in\mathbb{C}^{N_{\mathrm{I}}\times1}$
denote the phase-shift vector at the IRS, where $\theta_{l}=e^{j\phi_{l}}$,
$l\in\mathcal{N}_{\mathrm{I}}\triangleq\left\{ 1,2,\ldots,N_{\mathrm{I}}\right\} $,
and $\phi_{l}\in[0,2\pi)$. Here the ST, SR and IRS constitute the
secondary network, and in order to communicate with the SR, the ST
uses the spectrum band whose license is owned by a primary network.
However, with the help of properly designed beamformers at the ST
and IRS, it is ensured that the interference inflicted by the secondary
network on the PRs is below a predefined threshold.\footnote{Following the arguments in the related works of~\cite{Larsson_MISO_TCOM,Robust_MISO_TCCN,MIMO_BCD_TVT,Limeng_WCNC,ICT_Invited},
we assume that the secondary network is deployed outside the coverage
of the primary transmitter, and that the distance between the primary
transmitter and secondary receiver is large enough for the primary-to-secondary
interference to be negligible. \textcolor{black}{However, it is noteworthy
that the proposed PDDGP algorithm is also applicable when the secondary
network is deployed completely or partially inside the coverage of
the primary transmitter.}}

The signal received at the SR is given by
\begin{equation}
\begin{aligned} & \mathbf{y}_{\mathrm{R}}=(\mathrm{\mathbf{H}_{\mathrm{IR}}}\boldsymbol{\Theta}\mathbf{H}_{\mathrm{TI}}+\mathbf{H}_{\mathrm{TR}})\mathbf{x}+\mathbf{n}_{\mathrm{R}},\end{aligned}
\label{eq:RxSignals}
\end{equation}
where $\boldsymbol{\Theta}\triangleq\diag(\boldsymbol{\theta})$,
$\mathbf{x}\in\mathbb{C}^{N_{\mathrm{T}}\times1}$ is the transmitted
signal vector from the ST (which is intended for the SR), $\mathbf{n}_{\mathrm{R}}\sim\mathcal{CN}(\boldsymbol{0},\sigma^{2}\mathbf{I})$
denotes the complex additive white Gaussian noise (AWGN) vector at
the SR, and $\sigma^{2}$ is the noise power\textcolor{black}{. Without
loss of generality, we assume that the noise at all of the PRs is
also distributed as $\mathcal{CN}(\boldsymbol{0},\sigma^{2}\mathbf{I})$.
With a slight abuse of notation, in the rest of letter, we make the
replacements $\mathbf{H}_{\mathrm{TI}}\leftarrow\mathbf{H}_{\mathrm{TI}}/\sigma$,
$\mathbf{H}_{\mathrm{TR}}\leftarrow\mathbf{H}_{\mathrm{TR}}/\sigma$
and $\mathbf{H}_{\mathrm{T}k}\leftarrow\mathbf{H}_{\mathrm{T}k}/\sigma$.
This normalization step will also mitigate potential numerical issues
since we avoid dealing with extremely small quantities. }

\paragraph*{Problem Formulation}

Denote the transmit covariance matrix at the ST by $\mathbf{X}\triangleq\mathbb{E}\{\mathbf{x}\mathbf{x}\herm\}$,
the maximum transmit power budget at the ST by $P_{\max}$, and the
maximum tolerable interference power (normalized to the noise power)
at the $k$-th PR by $P_{k}$. Then, the achievable rate maximization
problem for the secondary network can be formulated as follows:

\begin{subequations}
\label{eq:MainProblem}

\begin{align}
\underset{\mathbf{\mathbf{X}\succeq}\boldsymbol{0},\boldsymbol{\theta}}{\maximize}\  & R(\mathbf{X},\boldsymbol{\theta})=\ln|\mathbf{I}+\mathbf{Z}_{\mathrm{R}}\mathbf{X}\mathbf{Z}_{\mathrm{R}}\herm|,\label{eq:OptProb_Obj}\\
\st\  & \tr(\mathbf{X})\leq P_{\max},\label{eq:SPC}\\
 & \tr(\mathbf{Z}_{k}\mathbf{X}\mathbf{Z}_{k}\herm)\leq P_{k},\forall k\in\mathcal{K},\label{eq:IPC}\\
 & |\theta_{l}|=1,\forall l\in\mathcal{N}_{\mathrm{I}},\label{eq:UMC}
\end{align}
\end{subequations}
whe\textcolor{black}{re $\mathbf{Z}_{\mathrm{R}}\triangleq\mathrm{\mathbf{H}_{\mathrm{IR}}}\boldsymbol{\Theta}\mathbf{H}_{\mathrm{TI}}+\mathbf{H}_{\mathrm{TR}}$,
and $\mathbf{Z}_{k}\triangleq\mathbf{H}_{\mathrm{I}k}\boldsymbol{\Theta}\mathbf{H}_{\mathrm{TI}}+\mathbf{H}_{\mathrm{T}k}$.
N}ote that the constraints in~\eqref{eq:SPC}--\eqref{eq:UMC} represent
the transmit power constraint at the ST, the interference power constraints
(IPCs) at the PRs and the unit-modulus constraints (UMCs) at the IRS,
respectively. Furthermore, \textcolor{black}{the feasible sets for
$\mathbf{X}$ and $\boldsymbol{\theta}$ are defined }as $\mathcal{X}\triangleq\{\mathbf{X}\in\mathbb{C}^{N_{\mathrm{T}}\times N_{\mathrm{T}}}:\mathrm{Tr}(\mathbf{X})\leq P_{\max};\mathbf{X}\succeq\boldsymbol{0}\}$
and $\vartheta\triangleq\{\boldsymbol{\theta}\in\mathbb{C}^{N_{\mathrm{I}}\times1}:\left|\theta_{l}\right|=1,l\in\mathcal{N}_{\mathrm{I}}\}$,
respectively. 

\section{\label{sec:Proposed-Solution}Proposed Solution}

\subsection{PDDGP Algorithm}

It is easy to see that problem~\eqref{eq:MainProblem} is non-convex
due to the non-convexity of the objective as well as constraints \eqref{eq:IPC}
and \eqref{eq:UMC}. If the IPCs were not present, \eqref{eq:MainProblem}
would reduce to the conventional MIMO capacity problem for which the
gradient projection algorithm in~\cite{Stefan_TWC} can be applied
to obtain a stationary solution. In fact, it is the coupling of $\mathbf{X}$
and $\boldsymbol{\theta}$ due to the IPCs in \eqref{eq:IPC} that
makes the problem much more difficult to solve and the gradient projection
algorithm proposed in~\cite{Stefan_TWC} is no longer applicable.
The main idea of the existing optimization methods in \cite{TWC_Ding,MIMO_BCD_TVT}
for solving \eqref{eq:MainProblem} is to alternately optimize $\mathbf{X}$
and $\boldsymbol{\theta}$. However, due to the coupling of $\mathbf{X}$
and $\boldsymbol{\theta}$, these methods are not guaranteed to return
a stationary solution. Indeed we numerically demonstrate in Section~\ref{sec:Results-and-Discussion}
that these methods are inferior to the PDDGP algorithm that we propose
in the following.

Specifically, in order to cope with the IPCs, we apply the penalty
dual decomposition (PDD) method recently presented in~\cite{Penalty_TSP}.
First, let $g_{k}(\mathbf{X},\boldsymbol{\theta},P_{k},s_{k})\triangleq\tr(\mathbf{Z}_{k}\mathbf{X}\mathbf{Z}_{k}\herm)+s_{k}-P_{k}$.
Then it is easy to see that~\eqref{eq:IPC} is equivalent to $g_{k}(\mathbf{X},\boldsymbol{\theta},P_{k},s_{k})=0,s_{k}\geq0,\forall k\in\mathcal{K}$.
Next, \textcolor{black}{the augmented Lagrangian function can be written
as follows:}
\begin{align*}
 & \hat{R}_{\boldsymbol{\upsilon},\rho}(\mathbf{X},\boldsymbol{\theta},\mathbf{s})\triangleq\ln|\mathbf{I}+\mathbf{Z}_{\mathrm{R}}\mathbf{X}\mathbf{Z}_{\mathrm{R}}\herm|\\
 & \qquad-\sum_{k\in\mathcal{K}}\upsilon_{k}g_{k}(\mathbf{X},\boldsymbol{\theta},P_{k},s_{k})-\frac{1}{2\rho}\sum_{k\in\mathcal{K}}g_{k}^{2}(\mathbf{X},\boldsymbol{\theta},P_{k},s_{k}),
\end{align*}
where $\mathbf{s}\triangleq[s_{1},s_{2},\ldots,s_{K}]\trans$, $\boldsymbol{\upsilon}\triangleq[\upsilon_{1},\upsilon_{2},\ldots,\upsilon_{K}]\trans$,
$\upsilon_{k}$ is the Lagrangian multiplier corresponding to the
constraint $g_{k}(\mathbf{X},\boldsymbol{\theta},P_{k},s_{k})=0$
and $\rho$ is the \textit{\emph{penalty parameter}}. Hence, for a
given $(\boldsymbol{\upsilon},\rho)$ \textcolor{black}{an equivalent
optimization problem can be formulated as }
\begin{subequations}
\label{eq:OptProbTransformed}
\begin{align}
\underset{\mathbf{X},\boldsymbol{\theta},\mathbf{s}}{\maximize} & \ \hat{R}_{\boldsymbol{\upsilon},\rho}(\mathbf{X},\boldsymbol{\theta},\mathbf{s}),\label{eq:TransformedObjective}\\
\st & \ \mathbf{s}\geq\boldsymbol{0},\eqref{eq:SPC},\eqref{eq:UMC}.\label{eq:PenaltyConstraint}
\end{align}
\end{subequations}
 Since the IPCs (and thus the coupling of $\mathbf{X}$ and $\boldsymbol{\theta}$)
is now included in the augmented objective, the optimization variables
in~\eqref{eq:OptProbTransformed} are decoupled. As a result, a simple
but efficient method \textcolor{black}{can be}\textcolor{blue}{{} }\textcolor{black}{applied}
to find a stationary solution to~\eqref{eq:OptProbTransformed} which
is based on the alternating projected gradient method (APGM). The
APGM is motivated by the fact that the feasible set with respect to
individual variables is simple in the sense that the projection onto
this set can be expressed in closed form.

First, \textcolor{black}{a projected gradient step for $\boldsymbol{\theta}$
is performed} while other variables are held fixed. In this regard,
the gradient of $\hat{R}_{\boldsymbol{\upsilon},\rho}(\mathbf{X},\boldsymbol{\theta},\mathbf{s})$
w.r.t. $\boldsymbol{\theta}$ is given by 
\begin{align}
 & \nabla_{\boldsymbol{\theta}}\hat{R}_{\boldsymbol{\upsilon},\rho}(\mathbf{X},\boldsymbol{\theta},\mathbf{s})=\nabla_{\boldsymbol{\theta}}\ln|\mathbf{I}+\mathbf{Z}_{\mathrm{R}}\mathbf{X}\mathbf{Z}_{\mathrm{R}}\herm|\nonumber \\
 & -\sum_{k\in\mathcal{K}}\big[\upsilon_{k}+\frac{1}{\rho}g_{k}(\mathbf{X},\boldsymbol{\theta},P_{k},s_{k})\big]\nabla_{\boldsymbol{\theta}}g_{k}(\mathbf{X},\boldsymbol{\theta},P_{k},s_{k}).\label{eq:gradThetaIncomplete}
\end{align}
Next using~\eqref{eq:gradThetaIncomplete},~\cite[eqn. (17a)]{Stefan_TWC}
and~\cite[Table 4.3 and eqn. (6.153)]{DerivativeBook}, $\nabla_{\boldsymbol{\theta}}\hat{R}_{\boldsymbol{\upsilon},\rho}(\mathbf{X},\boldsymbol{\theta},\mathbf{s})$
is given by~\eqref{eq:grad_theta} shown on the next page. Suppose
that, from the current point $\boldsymbol{\theta}$, after moving
along $\nabla_{\boldsymbol{\theta}}\hat{R}_{\boldsymbol{\upsilon},\rho}(\mathbf{X},\boldsymbol{\theta},\mathbf{s})$
with some step size \textcolor{black}{one arrives at a poi}nt $\hat{\boldsymbol{\theta}}=[\hat{\theta}_{1},\hat{\theta}_{2},\ldots,\hat{\theta}_{N_{\mathrm{I}}}]\trans\in\mathbb{C}^{N_{\mathrm{I}}\times1}$.
To obtain the next iterate, \textcolor{black}{the projection of $\hat{\boldsymbol{\theta}}$
onto the set $\vartheta$ is needed.} 
\begin{figure*}[t]
\begin{align}
 & \nabla_{\boldsymbol{\theta}}\hat{R}_{\boldsymbol{\upsilon},\rho}(\mathbf{X},\boldsymbol{\theta},\mathbf{s})\!=\!\vecd\big(\mathbf{H}_{\mathrm{IR}}\herm\big(\mathbf{I}+\mathbf{Z}_{\mathrm{R}}\mathbf{X}\mathbf{Z}_{\mathrm{R}}\herm\big)^{-1}\mathbf{Z}_{\mathrm{R}}\mathbf{X}\mathbf{H}_{\mathrm{TI}}\herm\big)\!-\!\sum_{k\in\mathcal{K}}\big[\upsilon_{k}\!+\!\tfrac{1}{\rho}\{\tr(\mathbf{Z}_{k}\mathbf{X}\mathbf{Z}_{k}\herm)\!+\!s_{k}\!-\!P_{k}\}\big]\vecd(\ensuremath{\mathbf{H}_{\mathrm{I}k}\herm\mathbf{Z}_{k}\mathbf{X}\mathbf{H}_{\mathrm{TI}}\herm}).\label{eq:grad_theta}
\end{align}
\end{figure*}
It is easy to see that this projection is given by $\Pi_{\vartheta}(\hat{\boldsymbol{\theta}})=[\bar{\theta}_{1},\bar{\theta}_{2},\ldots,\bar{\theta}_{N_{\mathrm{I}}}]\trans$,
where for each $l\in\mathcal{N}_{\mathrm{I}}$ 
\begin{equation}
\bar{\theta}_{l}=\left\{ \begin{array}{cc}
\hat{\theta}_{l}/\big|\hat{\theta}_{l}\big|,\qquad\qquad\qquad\quad\  & \text{if }\hat{\theta}_{l}\neq0\\
\exp(j\phi),\phi\in[0,2\pi), & \text{otherwise}.
\end{array}\right.\label{eq:Proj_theta}
\end{equation}
Next,\textcolor{black}{{} a projected gradient step for $\mathbf{X}$
is carried out. Foll}owing a similar line of argument and using~\cite[eqns. (6.207), (6.195) and Table 4.3]{DerivativeBook},
a closed-form expression for the gradient of $\hat{R}_{\boldsymbol{\upsilon},\rho}(\mathbf{X},\boldsymbol{\theta},\mathbf{s})$
w.r.t. $\mathbf{X}$ can be obtained in~\eqref{eq:grad_X}, shown
on the next page.
\begin{figure*}[t]
\vskip-0.2in
\begin{equation}
\nabla_{\mathbf{X}}\hat{R}_{\boldsymbol{\upsilon},\rho}(\mathbf{X},\boldsymbol{\theta},\mathbf{s})=\mathbf{Z}_{\mathrm{R}}\herm(\mathbf{I}+\mathbf{Z}_{\mathrm{R}}\mathbf{X}\mathbf{Z}_{\mathrm{R}}\herm)^{-1}\mathbf{Z}_{\mathrm{R}}-\sum_{k\in\mathcal{K}}\big[\upsilon_{k}+\tfrac{1}{\rho}\{\tr(\mathbf{Z}_{k}\mathbf{X}\mathbf{Z}_{k}\herm)+s_{k}-P_{k}\}\big]\mathbf{Z}_{k}\herm\mathbf{Z}_{k}.\label{eq:grad_X}
\end{equation}

\hrulefill
\end{figure*}
 Moreover, the projection of a given point $\mathbf{W}\in\mathbb{C}^{N_{\mathrm{T}}\times N_{\mathrm{T}}}$
onto the feasible set $\mathcal{X}$ can be shown to admit a water-filling
solution. Due to space constraints, \textcolor{black}{the details
regarding the projection of $\mathbf{W}$ onto $\mathcal{X}$ are
omitted, while the} interested reader may refer to~\cite[Sec. III-C]{Stefan_TWC}
for details.

\begin{algorithm}[t]
\caption{The PDDGP algorithm.}

\label{algoPDDGP}

\KwIn{ $\boldsymbol{\theta}_{0}$, $\mathbf{X}_{0}$, $\mathbf{s}_{0}$,
$\boldsymbol{\upsilon}_{0}$, $\mu_{0}$, $\alpha_{0}$, $\rho$,
$\kappa<1$ }

\KwOut{ $\boldsymbol{\theta}^{\star},\mathbf{X}^{\star}$}

\Repeat{convergence }{

Solve problem~\eqref{eq:OptProbTransformed} using \textbf{Algorithm}~\ref{algoGP}\;

$\mathbf{X}^{\star}\leftarrow\mathbf{X}_{n}$, $\boldsymbol{\theta}^{\star}\leftarrow\boldsymbol{\theta}_{n}$,
$\mathbf{s}^{\star}\leftarrow\mathbf{s}_{n}$\;

$\upsilon_{k}\leftarrow\upsilon_{k}+\frac{1}{\rho}g_{k}(\mathbf{X}^{\star},\boldsymbol{\theta}^{\star},P_{k},s_{k}^{\star})$,
$\forall k\in\mathcal{K}$\;

$\rho\leftarrow\kappa\rho$\;

}
\end{algorithm}

We now turn to the optimization over $\mathbf{s}$. It is easy to
see that the optimal solution to \eqref{eq:OptProbTransformed}, when
$\boldsymbol{\theta}$ and $\mathbf{X}$ are fixed, is simply given
by
\[
s_{n,k}=\max\{0,P_{k}\!-\!\tr(\mathbf{Z}_{k}\mathbf{X}_{n}\mathbf{Z}_{k}\herm)\},\forall k\in\mathcal{K}
\]
and thus a projection step for $\mathbf{s}$ is not necessary. \textcolor{black}{The
proposed APGM to find a stationary solution to~\eqref{eq:OptProbTransformed}
for fixed $\boldsymbol{\upsilon}$ and $\rho$ is summarized }in~\textbf{Algorithm}~\textbf{\ref{algoGP}}.
Note that in~\textbf{Algorithm}~\textbf{\ref{algoGP}}, $\mu_{n}$
and $\alpha_{n}$ denote the step size corresponding to the gradient
step for the $\boldsymbol{\theta}$ and $\mathbf{X}$ update in iteration
$n$, respectively. Appropriate values of $\mu_{n}$ and $\alpha_{n}$
can be obtained by a backtracking\emph{ }\textit{\emph{line search.
Specifically,}} we can set $\mu_{n}=\mu_{n-1}\gamma^{i_{n}}$ where
$i_{n}$ is the smallest positive integer such that
\begin{align}
 & \hat{R}_{\boldsymbol{\upsilon},\rho}(\mathbf{X}_{n-1},\boldsymbol{\theta}_{n},\mathbf{s}_{n-1})\geq\hat{R}_{\boldsymbol{\upsilon},\rho}(\mathbf{X}_{n-1},\boldsymbol{\theta}_{n-1},\mathbf{s}_{n-1})\nonumber \\
 & \qquad\qquad\qquad+\langle\nabla_{\boldsymbol{\theta}}\hat{R}_{\boldsymbol{\upsilon},\rho}(\mathbf{X}_{n-1},\boldsymbol{\theta}_{n-1},\mathbf{s}_{n}),\boldsymbol{\theta}_{n}\!-\!\boldsymbol{\theta}_{n-1}\rangle\nonumber \\
 & \qquad\qquad\qquad\qquad\qquad-\dfrac{1}{\mu_{n-1}\gamma^{i_{n}}}\left\Vert \boldsymbol{\theta}_{n}-\boldsymbol{\theta}_{n-1}\right\Vert ^{2},\label{eq:LineSearchTheta}
\end{align}
where $\langle\mathbf{x},\mathbf{y}\rangle\triangleq2\Re(\mathbf{x}\herm\mathbf{y})$
and $\gamma<1$. A similar routine can be followed to obtain $\alpha_{n}$.

After solving \eqref{eq:OptProbTransformed} for given $(\boldsymbol{\upsilon},\rho)$,
\textcolor{black}{the penalty parameter, $\rho$, is decreased and
the Lagrange multipliers are updated. }The overall description of
the proposed PDDGP algorithm to find a stationary solution to \eqref{eq:OptProbTransformed}
is outlined in \textbf{Algorithm~\ref{algoPDDGP}.}

\subsection{Extension to Multiple SRs}

We have presented the system model where there is a single SR for
the sake of simplicity. However, we remark that the extension to deal
with multiple SRs is straightforward. In this case, we can consider
the maximization of the weighted sum rate (WSR) of all SRs as in~\cite{MIMO_BCD_TVT,TWC_Ding}
and the transmit power constraint~\eqref{eq:SPC} becomes the sum
power constraint. It is easy to check that the projected gradient
step for $\boldsymbol{\theta}$ requires minimal modifications. Also,
the projection onto the set defined by the sum power constraint still
admits a water-filling solution. Thus, it is trivial to see that~\textbf{Algorithm~\ref{algoPDDGP}}
can be easily modified to solve the resulting WSR maximization problem.
The detailed derivations are skipped here due to space limitation
and \textcolor{black}{only the numerical results are presented fo}r
the multiple SR case in the next section.

\subsection{Proof of Convergence}

The convergence proof of \textbf{Algorithm~\ref{algoPDDGP}} follows
the same arguments as given in~\cite{Penalty_TSP}, and thus we only
summarize the main points here. First, for given $(\boldsymbol{\upsilon},\rho)$,
\textbf{Algorithm~\ref{algoGP}} generates a strictly increasing
objective sequence, $\hat{R}_{\boldsymbol{\upsilon},\rho}(\mathbf{X},\boldsymbol{\theta},\mathbf{s})$.
Since the feasible set is bounded, the iterates returned by \textbf{Algorithm~\ref{algoGP}
}converge to a limit point of~\eqref{eq:TransformedObjective} to
some accuracy $\epsilon_{k}$. Next, it can be shown that the sequence
$\left\Vert \boldsymbol{\upsilon}_{k+1}-\boldsymbol{\upsilon}_{k}\right\Vert $
is bounded. Thus, from the dual update it holds that $\sqrt{\sum_{k\in\mathcal{K}}g_{k}^{2}(\mathbf{X},\boldsymbol{\theta},P_{k},s_{k})}=\rho\left\Vert \boldsymbol{\upsilon}_{k+1}-\boldsymbol{\upsilon}_{k}\right\Vert \to0$
as $\rho\to0$.

\subsection{Complexity Analysis}

\begin{figure*}[t]
\begin{minipage}[t]{0.24\textwidth}%
\centering
\includegraphics[width=0.9\columnwidth]{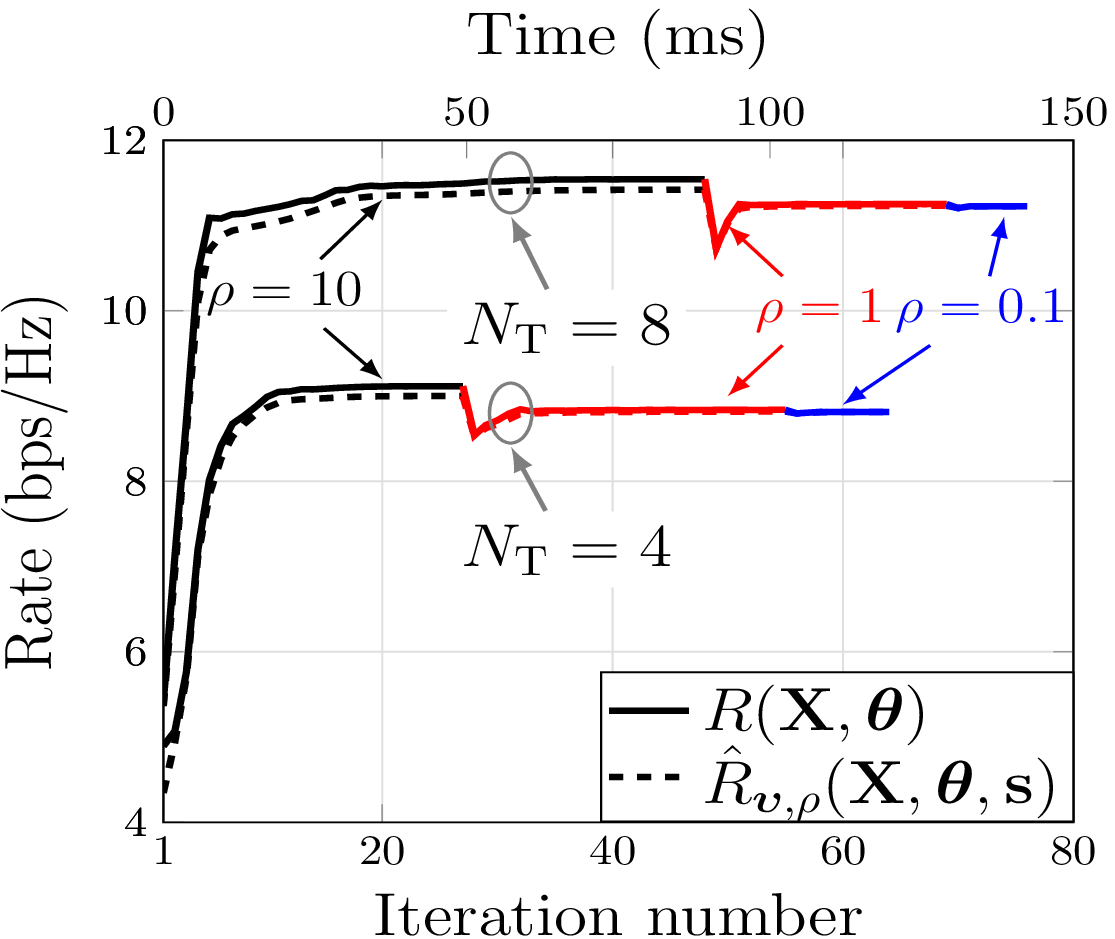}
\vspace{-0.2cm}\caption{Convergence results for $P_{\max}=20$ dBm.}
\label{fig:Convergence}%
\end{minipage}\hfill{}%
\begin{minipage}[t]{0.24\textwidth}%
\centering
\includegraphics[width=0.9\columnwidth]{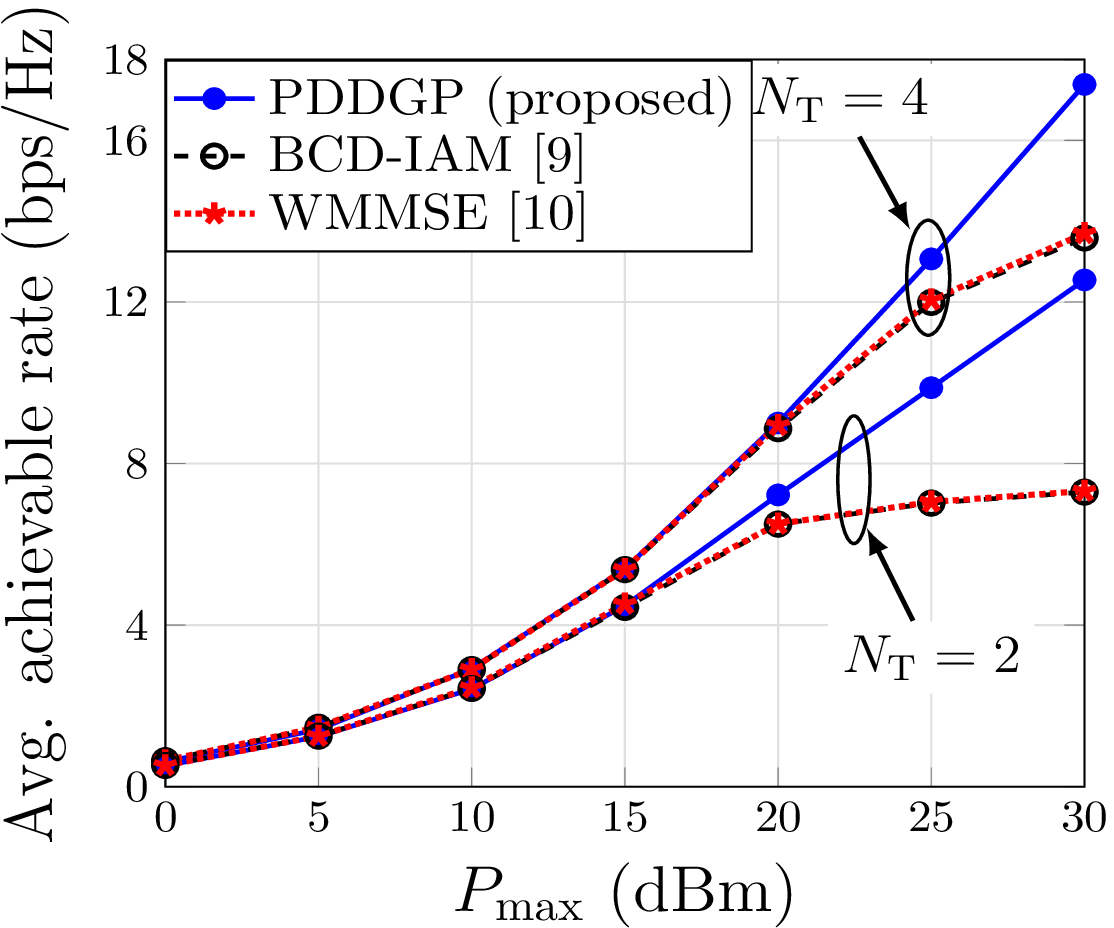}
\vspace{-0.2cm}\caption{Average achievable rate for a single PR.}
\label{fig:PDD_BCD_Comp}%
\end{minipage}\hfill{}%
\begin{minipage}[t]{0.24\textwidth}%
\centering
\includegraphics[width=0.9\columnwidth]{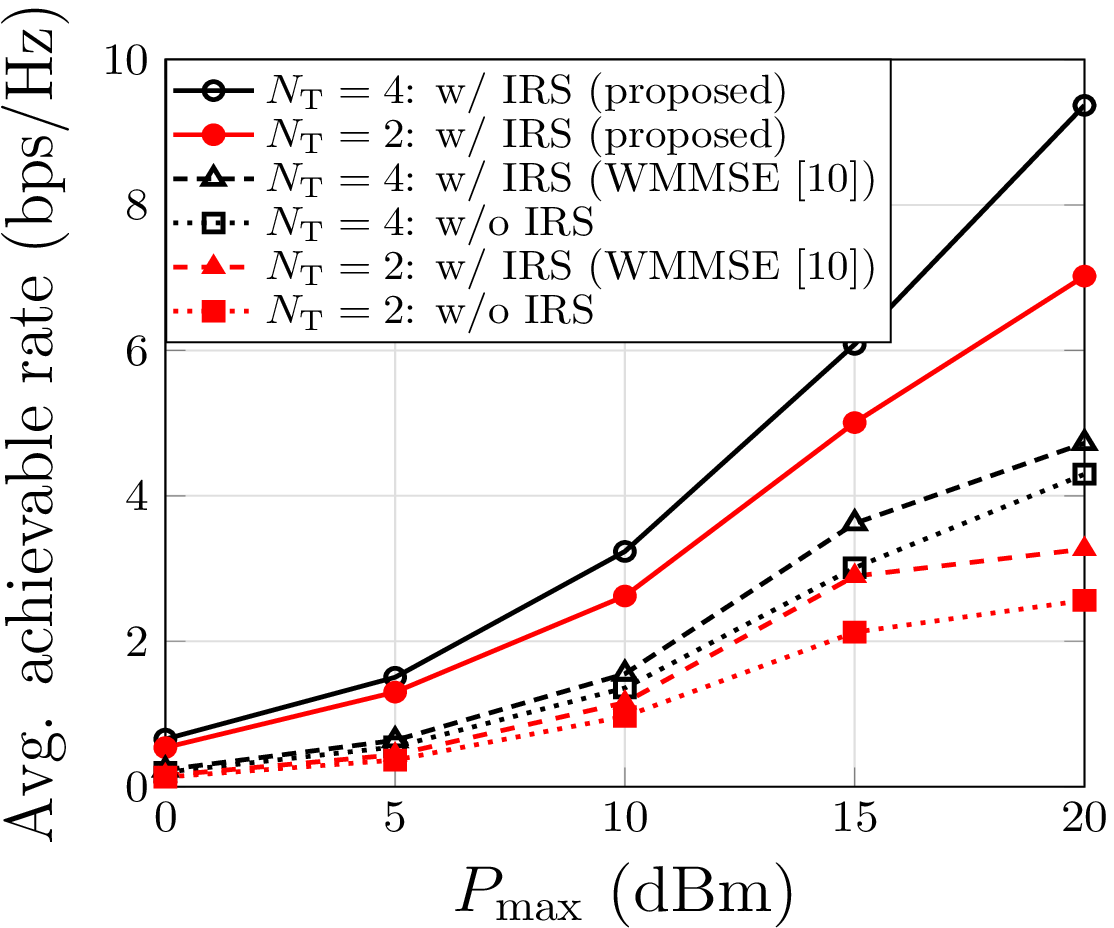}
\vspace{-0.2cm}\caption{Average achievable sum rate versus $P_{\max}$.}
\label{fig:Rate_vs_Pmax}%
\end{minipage}\hfill{}%
\begin{minipage}[t]{0.24\textwidth}%
\centering
\includegraphics[width=0.9\columnwidth]{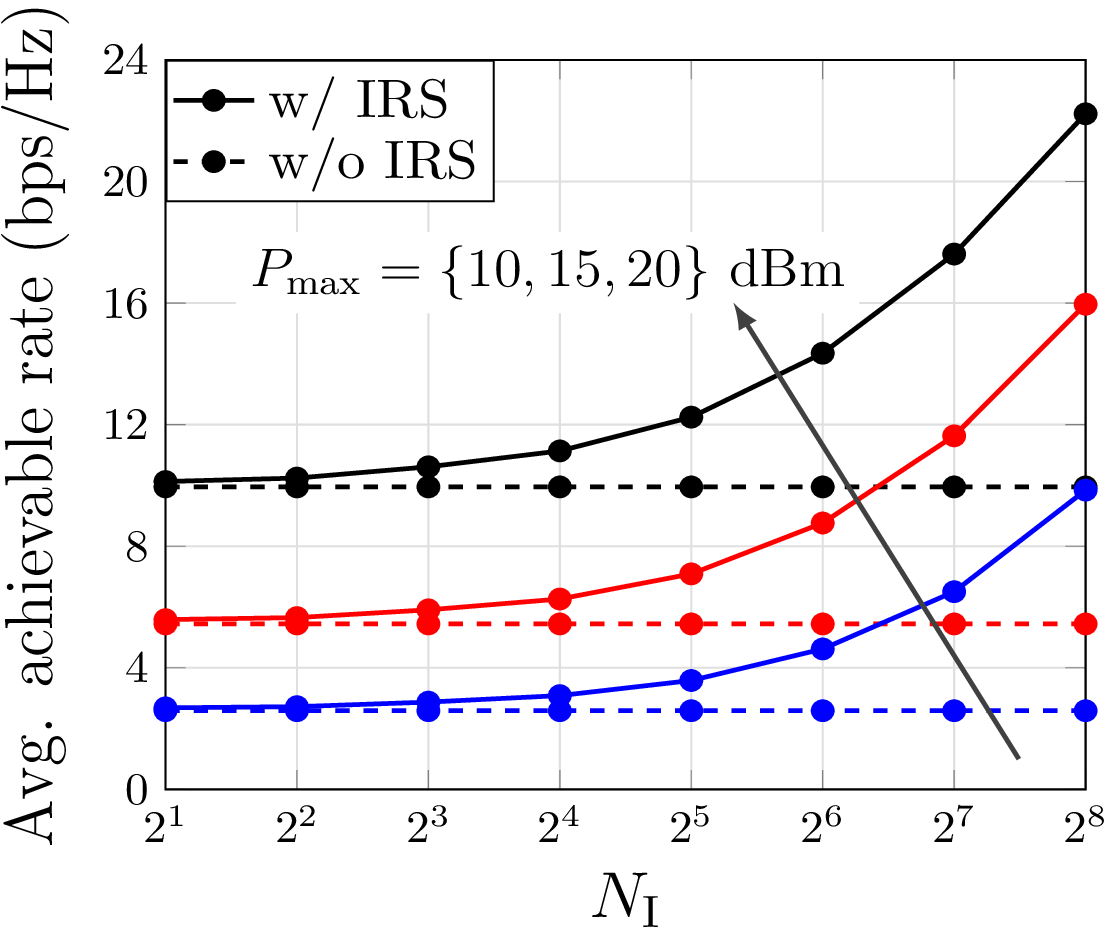}
\vspace{-0.2cm}\caption{Average achievable rate versus $N_{\mathrm{I}}$.}
\label{fig:Rate_vs_NI}%
\end{minipage}
\end{figure*}
It is obvious that the complexity of the proposed PDDGP algorithm
is dominated by that of \textbf{Algorithm \ref{algoGP}}. Thus\textcolor{black}{{}
the complexity of }\textbf{\textcolor{black}{Algorithm \ref{algoGP}}}\textcolor{black}{{}
is obtained} by counting the required number of complex multiplications
using the big-O notation. First, the computational complexity associated
with $\nabla_{\boldsymbol{\theta}}\hat{R}_{\boldsymbol{\upsilon},\rho}(\mathbf{X},\boldsymbol{\theta},\mathbf{s})$
in~\eqref{eq:grad_theta} is as follows. The complexity for computing
the terms $\mathbf{Z}_{\mathrm{R}}$, $\mathbf{Z}_{\mathrm{R}}\mathbf{X}\mathbf{Z}_{\mathrm{R}}\herm$,
$(\mathbf{I}+\mathbf{Z}_{\mathrm{R}}\mathbf{X}\mathbf{Z}_{\mathrm{R}}\herm)^{-1}\mathbf{Z}_{\mathrm{R}}\ (\triangleq\boldsymbol{\Xi})$,
$\boldsymbol{\Xi}\mathbf{X}$ and $\vecd(\hat{\mathbf{H}}_{\mathrm{IR}}\herm\boldsymbol{\Xi}\mathbf{X}\mathbf{H}_{\mathrm{TI}}\herm)$
is respectively $O(N_{\mathrm{R}}N_{\mathrm{I}}N_{\mathrm{T}})$,
$O(N_{\mathrm{R}}N_{\mathrm{T}}^{2}+N_{\mathrm{R}}^{2}N_{\mathrm{T}})$,
$O(N_{\mathrm{R}}^{3}+N_{\mathrm{R}}^{2}N_{\mathrm{T}})$, $O(N_{\mathrm{R}}N_{\mathrm{T}}^{2})$
and $O(N_{\mathrm{I}}N_{\mathrm{T}}N_{\mathrm{R}})$. Similarly, the
complexity for obtaining $\tr(\mathbf{Z}_{k}\mathbf{X}\mathbf{Z}_{k}\herm)$
\textcolor{black}{and $\vecd(\mathbf{H}_{\mathrm{I}k}\herm\mathbf{Z}_{k}\mathbf{X}\mathbf{H}_{\mathrm{TI}}\herm)$
is respectively $O(N_{\mathrm{P}}N_{\mathrm{T}}(N_{\mathrm{I}}+N_{\mathrm{T}}))$
and $O(N_{\mathrm{T}}^{2}N_{\mathrm{I}}+N_{\mathrm{P}}N_{\mathrm{T}}N_{\mathrm{I}})$.
Therefore, the total complexity for calculating $\nabla_{\boldsymbol{\theta}}\hat{R}_{\boldsymbol{\upsilon},\rho}(\mathbf{X},\boldsymbol{\theta},\mathbf{s})$
is $O(2N_{\mathrm{R}}N_{\mathrm{I}}N_{\mathrm{T}}+2N_{\mathrm{R}}N_{\mathrm{T}}^{2}+2N_{\mathrm{R}}^{2}N_{\mathrm{T}}+N_{\mathrm{R}}^{3}+(K+1)N_{\mathrm{P}}N_{\mathrm{T}}N_{\mathrm{I}}+KN_{\mathrm{T}}^{2}N_{\mathrm{I}}+KN_{\mathrm{P}}N_{\mathrm{T}}^{2}).$
We remark that, since the projection onto $\vartheta$ has negligible
complexity, the additional complexity to find an appropriate value
of $\mu_{n}$ using the backtracking line search in \eqref{eq:LineSearchTheta}
can be ignored.}

\textcolor{black}{Next, the complexity to compute $\nabla_{\mathbf{X}}\hat{R}_{\boldsymbol{\upsilon},\rho}(\mathbf{X},\boldsymbol{\theta},\mathbf{s})$
is $O(N_{\mathrm{T}}^{2}(N_{\mathrm{R}}+KN_{\mathrm{P}}))$. Additionally,
the complexity for projecting $\mathbf{X}_{n\!-\!1}+\alpha_{n\!-\!1}\nabla_{\mathbf{X}}\hat{R}_{\boldsymbol{\upsilon},\rho}(\mathbf{X}_{n\!-\!1},\boldsymbol{\theta}_{n},\mathbf{s}_{n\!-\!1})$
onto the feasible set $\mathcal{X}$ is $O(N_{\mathrm{T}}^{3})$.
Therefore, the computational complexity for the projected gradient
step for $\mathbf{X}$ is $O(N_{\mathrm{R}}N_{\mathrm{T}}^{2}+KN_{\mathrm{P}}N_{\mathrm{T}}^{2}+N_{\mathrm{T}}^{3}).$}

Moreover, it is straightforward to observe that computing $\mathbf{s}_{n}$
requires $O(K(N_{\mathrm{P}}N_{\mathrm{I}}N_{\mathrm{T}}+N_{\mathrm{P}}N_{\mathrm{T}}^{2}))$
complex multiplications. Therefore, the per-iteration complexity of~\textbf{Algorithm~\ref{algoGP}}
\textcolor{black}{is $O(N_{\mathrm{R}}^{3}+N_{\mathrm{T}}^{3}+2N_{\mathrm{R}}^{2}N_{\mathrm{T}}+N_{\mathrm{T}}^{2}(3N_{\mathrm{R}}+KN_{\mathrm{I}}+3KN_{\mathrm{P}})+N_{\mathrm{T}}N_{\mathrm{I}}(2N_{\mathrm{R}}+2KN_{\mathrm{P}})).$}

It is noteworthy that in a practical IRS-assisted MIMO underlay spectrum
sharing system, $N_{\mathrm{I}}\gg\max\{N_{\mathrm{T}},N_{\mathrm{R}},N_{\mathrm{P}}\}$,
and therefore, the per-iteration complexity of~\textbf{Algorithm~\ref{algoGP}}
can be approximated\textcolor{black}{{} by $O(N_{\mathrm{I}}(KN_{\mathrm{T}}^{2}+2N_{\mathrm{R}}N_{\mathrm{T}}+2KN_{\mathrm{P}}))$,
which is linear w.r.t. $N_{\mathrm{I}}$. Note that}\textcolor{blue}{{}
}the complexity of the algorithms proposed in both~\cite{MIMO_BCD_TVT}
and~\cite{TWC_Ding} grows as $O(N_{\mathrm{I}}^{3}+KN_{\mathrm{I}}^{2}\max\{N_{\mathrm{P}},N_{\mathrm{T}}\})$
which is significantly higher than that of our proposed method.

\section{Results and Discussion\label{sec:Results-and-Discussion}}

In this section, \textcolor{black}{the results of numerical experiments
are presented to} evaluate the performance of the system under consideration.
We consider $K=4$ PRs (unless stated otherwise), and the location
of the ST, SR, IRS and the four PRs in Cartesian coordinates are respectively
given by $(300\,\mathrm{m},0\,\mathrm{m})$, $(600\,\mathrm{m},0\,\mathrm{m})$,
$(300\,\mathrm{m},30\,\mathrm{m})$ and $\{(0\,\mathrm{m},0\,\mathrm{m}),\,(0\,\mathrm{m},5\,\mathrm{m}),\,(0\,\mathrm{m},10\,\mathrm{m}),\,(0\,\mathrm{m},15\,\mathrm{m})\}$.
\textcolor{black}{The small-scale fading is assumed to be }Rayleigh
distributed, and that the path loss between two nodes is modeled as
$\mathrm{PL}=(-30-10\xi\log_{10}(d/d_{0}))$ dB, where $\xi$ is the
path loss exponent, $d$ is the distance between the nodes, and $d_{0}$
$(=1\,\mathrm{m})$ is the reference distance. We assume that the
path loss exponent for the ST-SR and ST-PR links is 3.75, whereas
that for the ST-IRS, IRS-SR and IRS-PR links is 2.2~\cite{MIMO_BCD_TVT}.
Furthermore, \textcolor{black}{it is assumed that }$N_{\mathrm{R}}=4$,
$N_{\mathrm{P}}=4$, $P_{k}=10^{-13}$~W $(\forall k\in\mathcal{K})$,
the noise power spectral density $\mathscr{N}_{0}=-174$~dBm/Hz and
the system bandwidth is $B=10$~MHz. In Figs.~\ref{fig:PDD_BCD_Comp}-\ref{fig:Rate_vs_NI},
the average achievable rate is computed over $100$ channel realizations.
To run \textbf{Algorithm}~\textbf{\ref{algoPDDGP}}, we start with
$\rho=10$ and decrease it as $\rho\leftarrow\kappa\rho$ where $\kappa=0.1$
when the relative objective progress of $\hat{R}_{\boldsymbol{\upsilon},\rho}(\mathbf{X},\boldsymbol{\theta},\mathbf{s})$
\cite[Eqn. (48)]{Penalty_TSP} in \textbf{Algorithm}~\textbf{\ref{algoGP}}
is less than $10^{-5}$. For \textbf{Algorithm}~\textbf{\ref{algoGP}}
we set $\gamma=0.5$ in the backtracking line search. \textbf{Algorithm}~\textbf{\ref{algoPDDGP}}
terminates when the difference between $R(\mathbf{X},\boldsymbol{\theta})$
and $\hat{R}_{\boldsymbol{\upsilon},\rho}(\mathbf{X},\boldsymbol{\theta},\mathbf{s})$
is less than $10^{-5}$.

In Fig.~\ref{fig:Convergence}, the convergence of \textbf{Algorithms}~\textbf{\ref{algoGP}}
and~\textbf{\ref{algoPDDGP}} \textcolor{black}{are shown for $N_{\mathrm{I}}=64$
and d}ifferent values of $N_{\mathrm{T}}$. In the figure, each iteration
is comprised of one update of $\mathrm{\mathbf{X}}$, $\boldsymbol{\theta}$
and $\mathrm{\mathbf{s}}$. Each color in Fig.~\ref{fig:Convergence}
in fact represents the convergence of \textbf{Algorithm}~\textbf{\ref{algoGP}}
for a fixed $\rho$ whose value is clearly indicated in the figure.
We recall that \textbf{Algorithm}~\textbf{\ref{algoGP}} aims to
maximize the augmented objective $\hat{R}_{\boldsymbol{\upsilon},\rho}(\mathbf{X},\boldsymbol{\theta},\mathbf{s})$,
and thus, for a given $\rho$, monotonic increase is only guaranteed
for $\hat{R}_{\boldsymbol{\upsilon},\rho}(\mathbf{X},\boldsymbol{\theta},\mathbf{s})$,
but not for the original objective $R(\mathbf{X},\boldsymbol{\theta})$.
This point is clearly seen in Fig.~\ref{fig:Convergence}. We also
observe that as the number of iterations increases, the difference
between $R(\mathbf{X},\boldsymbol{\theta})$ and the augmented objective
$\hat{R}_{\boldsymbol{\upsilon},\rho}(\mathbf{X},\boldsymbol{\theta},\mathbf{s})$
decreases and approaches zero for sufficiently small $\rho$, in accordance
with the principle of a penalty method. In the figure, we also show
the run time for \textbf{Algorithm}~\textbf{\ref{algoPDDGP}}, where
the algorithm converges before 150~ms for $N_{\mathrm{T}}=8$. Note
that we have implemented the algorithm using Python~v3.9.7 on a Linux
PC with 7.5 GiB memory and Intel Core i5-7200U CPU.

In Fig.~\ref{fig:PDD_BCD_Comp}, we demonstrate the superiority of
the proposed PDDGP algorithm over the BCD-IAM and WMMSE approaches
proposed in~\cite{MIMO_BCD_TVT} and~\cite{TWC_Ding}, respectively,
for $N_{\mathrm{I}}=64$. Note that here only the first PR located
at $(0\,\mathrm{m},0\,\mathrm{m})$ is considered since the algorithm
in~\cite{MIMO_BCD_TVT} was proposed for a single PR. It is evident
from the figure that all the schemes result in a similar performance
for small values of $P_{\max}$; however, the proposed PDDGP algorithm
offers a significantly higher average achievable rate for large values
of $P_{\max}$ compared to that achieved via the solutions proposed
in~\cite{MIMO_BCD_TVT} and~\cite{TWC_Ding}. Also, it is expected
that the SU rate performance achieved via the BCD-IAM and WMMSE schemes
is the same because both of these solutions are derived from similar
principles. The main reason for the inferior performance of the BCD-IAM
and WMMSE for large $P_{\max}$ is that as $P_{\max}$ increases,
the SU rate performance is mostly dominated by the IPCs at the PRs.
In other words, the IPCs become binding for large $P_{\max}$. Consequently,
the coupling between $\mathbf{X}$ and $\boldsymbol{\theta}$ becomes
stronger, and it is known that AO-based optimization cannot guarantee
a stationary solution under such a scenario.

Next, in Fig.~\ref{fig:Rate_vs_Pmax}, the average achievable sum
rate versus the maximum transmit power budget at the ST ($P_{\max}$)
\textcolor{black}{is shown for $N_{\mathrm{I}}=64$. In }particular,
we compare the proposed PDDGP algorithm to the WMMSE-based AO algorithm
introduced in~\cite{TWC_Ding} for the case of two SRs located at
$(600\,\mathrm{m},0\,\mathrm{m})$ and $(600\,\mathrm{m},5\,\mathrm{m})$.
In Fig.~\ref{fig:Rate_vs_Pmax}, it can be clearly seen that the
proposed PDDGP always outperforms the WMMSE-based algorithm. Note
that the BCD-IAM and the WMMSE-based AO are similar in principle and
thus share the same drawbacks. This explains why the WMMSE-based AO
in~\cite{TWC_Ding} is inferior to our proposed algorithm. However,
note that different from the case of a single PR in Fig.~\ref{fig:PDD_BCD_Comp},
the PDDGP algorithm outperforms the WMMSE algorithm even in the lower
$P_{\max}$ regime for the multiple PRs case, and the gains are more
significant for larger $P_{\max}$. The benefit of an IRS-assisted
MIMO system over that of the system without IRS can also be observed
from Fig. \ref{fig:Rate_vs_Pmax}, as the former results in a considerably
higher achievable rate.

\textcolor{black}{To investigate the effect of the number of reflecting
elements at the IRS, in Fig.~\ref{fig:Rate_vs_NI} we plot the average
achievable rate achieved by the proposed PDDGP algorithm for one SR
as a function of $N_{\mathrm{I}}$ for $N_{\mathrm{T}}=16$, $N_{\mathrm{R}}=4$,
$N_{\mathrm{P}}=2$ and $K=4$. Interestingly, for an exponential
increase in $N_{\mathrm{I}}$, the achievable rate of the MIMO system
also increases near-exponentially. This occurs because for large values
of $N_{\mathrm{I}}$, the IRS can create a highly-focused beam to
increase the signal-to-noise ratio (SNR) at the SR while satisfying
the interference constraints at the PRs, resulting in a higher achievable
rate at the SR.}

\section{Conclusion}

In this letter, we have presented the optimization problem for achievable
rate maximization in an IRS-assisted SU MIMO system in underlay spectrum
sharing with multiple PRs, subject to a transmit power constraint
at the ST, unit modulus constraints at the IRS and multiple interference
power constraints at the PRs. A PDDGP algorithm \textcolor{black}{has
been proposed t}o jointly optimize the transmit beamforming and the
IRS phase shifts. The numerical results established the superiority
of the proposed algorithm over two existing approaches in terms of
the achievable rate. We also showed that the complexity of the proposed
PDDGP algorithm is considerably lower than that of the existing approaches.

\bibliographystyle{ieeetr}
\bibliography{Kumar_WCL2022-0676}

\end{document}